\documentclass[letterpaper,10pt,svgnames]{article}
\usepackage[utf8]{inputenc}

\newcommand{\Rsym}{\ensuremath{\text{R}_{\text{sym}}}\xspace}

\newcommand{\y}{\ensuremath{y}\xspace}

\newcommand{\Ptx}{\ensuremath{P_{\text{tx}}}\xspace}

\newcommand{\SNR}{\ensuremath{\text{SNR}}}

\usepackage{color}
\usepackage{xcolor}
\usepackage{graphicx}
\usepackage{tikzscale}
\usepackage{tikz}
\usepackage{pgfplots}
\usetikzlibrary{plotmarks,matrix,chains,scopes,fit,calc,shapes,positioning,decorations,intersections,fit,backgrounds,patterns}
\usetikzlibrary{fadings,shapes.arrows,shadows}
\pgfplotsset{compat=newest} 

\pgfplotsset{plot coordinates/math parser=false}
\pgfplotsset{every axis plot/.append style={solid,line width=1.5pt,mark size=1.5pt,mark options={solid,fill=white}}}
\pgfplotsset{every axis legend/.append style={legend cell align=left,font=\footnotesize}}

\colorlet{42GBd16QAM_color}{blue!80!white}
\colorlet{42GBd64QAM_color}{red!80!black}
\colorlet{64GBd16QAM_color}{orange!90!black}
\colorlet{64GBd64QAM_color}{green!50!black}

\pgfplotsset{64GBd64QAM/.style={color=64GBd64QAM_color,solid}}
\pgfplotsset{64GBd16QAM/.style={color=64GBd16QAM_color,dotted}}
\pgfplotsset{42GBd64QAM/.style={color=42GBd64QAM_color,dashed}}
\pgfplotsset{42GBd16QAM/.style={color=42GBd16QAM_color,dash dot dot}}

\newlength\FigureWidth
\newlength\FigureHeight
\newlength\FullFigureWidth
\graphicspath{{figures/}}

\pgfplotsset{myLegend/.append style={legend style={font=\footnotesize,at={(0.5,0.98)},anchor=north,align=left,legend columns=3}}}

\usepackage{xparse}
\tikzfading[name=arrowfading, top color=transparent!0, bottom color=transparent!95]
\tikzset{arrowfill/.style={#1,general shadow={fill=black, shadow yshift=-0.8ex, path fading=arrowfading}}}
\tikzset{arrowstyle/.style n args={3}{draw=#2,arrowfill={#3}, single arrow,minimum height=#1, single arrow,
single arrow head extend=.3cm,}}

\NewDocumentCommand{\tikzfancyarrow}{O{2cm} O{FireBrick} O{top color=OrangeRed!20, bottom color=Red} m}{
\tikz[baseline=-0.5ex]\node [arrowstyle={#1}{#2}{#3}] {#4};
} 

\makeatletter
\tikzset{
    block filldraw/.style={
        draw, fill=yellow!20},
    block rect/.style={
        block filldraw, rectangle},
    block/.style={
        block rect, minimum height=0.8cm, minimum width=6em},
    from/.style args={#1 to #2}{
        above right={0cm of #1},
        /utils/exec=\pgfpointdiff
            {\tikz@scan@one@point\pgfutil@firstofone(#1)\relax}
            {\tikz@scan@one@point\pgfutil@firstofone(#2)\relax},
        minimum width/.expanded=\the\pgf@x,
        minimum height/.expanded=\the\pgf@y}}
\makeatother

\usepackage{osameet3} 
\usepackage{amsmath,amssymb}
\usepackage[normalem]{ulem}

\usepackage{xspace}
\usepackage{wrapfig}
\usepackage[tableposition=top]{caption}
\usepackage{enumitem}
\setlist[2]{noitemsep} 
\setenumerate{noitemsep}

\usepackage[colorlinks=true,bookmarks=false,citecolor=blue,urlcolor=blue]{hyperref} 
\usepackage{graphicx}
\graphicspath{{./figures/}}

\usepackage{floatrow}
\newfloatcommand{capbtabbox}{table}[][\FBwidth]
\begin{document}

\title{Estimating Quality of Transmission in a Live Production Network using Machine Learning}

\author{Jasper Müller\textsuperscript{(1)}, Tobias Fehenberger\textsuperscript{(1)}, Sai Kireet Patri\textsuperscript{(1)}, Kaida Kaeval\textsuperscript{(1)},\\ Helmut Griesser\textsuperscript{(1)}, Marko Tikas\textsuperscript{(2)}, and Jörg-Peter Elbers\textsuperscript{(1)}}
\address{\textsuperscript{(1)}ADVA, Fraunhoferstr. 9a, 82152 Martinsried/Munich, Germany
\textsuperscript{(2)}Tele2 Estonia, Tallinn, Estonia\\}
\vspace{-3pt}
\email{\href{mailto:jmueller@advaoptical.com}{jmueller@adva.com}}
\vspace{-14pt}
\begin{abstract} 
We demonstrate QoT estimation in a live network utilizing neural networks trained on synthetic data spanning a large parameter space. The ML-model predicts the measured lightpath performance with \textless0.5dB SNR error over a wide configuration range.\\
\end{abstract}

\vspace{-2pt}
\section{Introduction}
\vspace{-7pt}

Due to fast growing demand in optical network capacity, efficient usage of the available network infrastructure and therefore physical-layer aware optimization of the network throughput becomes increasingly important. An accurate method for estimating the quality of transmission (QoT) of unestablished lightpaths can be utilized to maximize transmission capacity. For the optimization of channel configurations, a fast method for QoT estimation is required.
Determining the QoT of a lightpath entails computing the linear noise contributions and the nonlinear interference (NLI), for which detailed and accurate knowledge on the system and network components is required. While linear noise calculation is relatively straightforward, the NLI computation can be a challenging task. Several models are available for this, such as numerical split-step simulations and Gaussian noise (GN) models \cite{EGN}, all following the general trade-off between complexity and accuracy.

In recent years, machine learning (ML) models for QoT estimation of unestablished lightpaths have been explored, using mainly synthetic data generated in simulations to leverage ML models for fast and accurate QoT estimation. In \cite{Morais}, a comparison of multiple models for predicting signal-to-noise ratio (SNR) margins is carried out considering fully loaded links only. Multiple classification models have been shown to achieve high accuracy in predicting whether the bit-error rate (BER) meets the threshold using a simplified parameter space, such as neural networks and support vector machines (SVM), assuming homogeneous fiber spans \cite{Aladin}, or random-forest algorithms \cite{Rottondi}, only considering the closest neighboring channels of a lightpath. In \cite{Gao}, the Q-value of multiple channels is predicted simultaneously with an artificial neural network (ANN) on a single testbed link collecting data from the link for training and verification. Such an ANN is a well suited ML technique for this task because it is able to learn highly nonlinear relationships between the input parameters. Previous work \cite{Morais}\cite{Aladin} as well as our own test show that an ANN is able to outperform other ML algorithms when used for QoT estimation. 

In this paper, we show, to the best of our knowledge, the first ANN-based QoT estimation of unestablished lightpaths in a live network with production channels. We generate synthetic network data, suitable to train an ML model that can accurately estimate the NLI for a large variety of link parameters and grid configurations. An input parameter space of significantly higher dimensionality is used than what is shown previously, which demands more complex ML model architectures. We demonstrate an ML model trained on the synthetic data that shows excellent prediction performance when applied to field measurements pulled from a live network. A maximum SNR deviation of less than 0.5~dB and an average SNR difference of less than 0.2~dB is achieved while computing the QoT of a single lightpath in microseconds, which is orders of magnitude faster than full-form GN models. 

\vspace{-9pt}
\section{Versatile ML Model based on Synthesized Simulation Data}
\vspace{-7pt}
\subsection{Simulation-based Data Generation}
\vspace{-5pt}
An ML model that works for a large variety of system and link configurations usually requires a big data set covering a large parameter space for training. Gathering this data in a live network is generally not possible because the variety of parameters available in a network is limited by the number of its lightpaths and traffic cannot be disrupted. We therefore resort to numerical simulations for generating the necessary data for training and then apply the trained model to the network under consideration. Due to the large required size of the data set and the large considered bandwidth, the EGN model \cite{EGN} was used. It is capable of accurately computing NLI and allowed creating a large database of $130,000$ data points in a reasonable amount of time. We used the nonlinear coefficient $\eta$ as output, which is related to SNR as $\SNR=\Ptx/(\sigma^2 + \eta P^3_\text{tx})$, where $\Ptx$ is the transmit power and the linear noise $\sigma^2$ is computed offline with the well-known closed-form expressions.

\begin{figure}\TopFloatBoxes
\begin{floatrow}
\ffigbox[0.55\textwidth]{%
  \tikzset{%
  every neuron/.style={
    circle,
    draw,
    minimum size=0.4cm
  },
  neuron missing/.style={
    draw=none, 
    scale=2.5,
    text height=0.333cm,
    execute at begin node=\color{black}$\vdots$
  },
}

\def\layersep{2.5cm}
\begin{tikzpicture}[x=0.65cm, y=0.65cm, >=stealth,font=\small]

 \begin{scope}[shift={(-0,0)}]
\foreach \m/\l [count=\y] in {1,2,missing,3}
  \node [every neuron/.try, neuron \m/.try] (input-\m) at (0,1.25-\y) {};

\foreach \m [count=\y] in {1,2,3,missing,4}
  \node [every neuron/.try, neuron \m/.try ] (hidden1-\m) at (2,2.5-\y*1.1) {};

\foreach \m [count=\y] in {1,2,3,missing,4}
  \node [every neuron/.try, neuron \m/.try ] (hidden2-\m) at (5,2.5-\y*1.1) {};

\foreach \m [count=\y] in {1}
  \node [every neuron/.try, neuron \m/.try ] (output-\m) at (7,0-\y/1.25) {};

\foreach \i in {1,...,3}
  \foreach \j in {1,...,4}
    \draw [->] (input-\i) -- (hidden1-\j);

\foreach \i in {1,...,4}
  \foreach \j in {1,...,4}
    \draw [->] (hidden1-\i) -- (hidden2-\j);

    \foreach \i in {1,...,4}
    \foreach \j in {1}
    \draw [->] (hidden2-\i) -- (output-\j);

\node [align=center, above] at (0,1.8) {\textbf{Input}\\\textbf{layer}};
\node [align=center, above] at (2,1.8) {\textbf{Hidden}\\\textbf{layer 1}};
\node [align=center, above] at (5,1.8) {\textbf{Hidden}\\\textbf{layer $n$}};
\node [align=center, above] at (7,1.8) {\textbf{Output}\\\textbf{layer}};

 \end{scope}

\node [align=center, above] at (-3.3,1.8) {\textbf{Input}\\\textbf{features}};
\node [align=left,text width=3.2cm,anchor=center,inner sep=0pt] (input_params) at (-3.3,-0.6) 
{
    \begin{itemize}[noitemsep,leftmargin=1mm]
        \item $\Ptx$ CUT
        \item $\Rsym$ CUT
        \item $4\times\{\Ptx$, Rsym, $\Delta f\}$
        \item Total used bandwidth
        \item \# of WDM channels
        \item \# of spans
        \item Span length parameters
        \item Average power level 
        \item $N_\text{ch}$ around CUT
    \end{itemize}
};
\draw[ultra thick,->] (-1.2,-0.8) -- ++(0.7,0);

\draw[ultra thick,->] (7.5,-0.8) -- ++(0.7,-0) node[right] {$\eta$};

\node[fit=(input-1)(hidden1-1)(hidden1-4)(output-1),draw,dashed,inner sep=1pt,rounded corners] {};
\end{tikzpicture}

}{%
  \vspace*{-\baselineskip}
  \caption{Schematic of the employed ML models.}
  \label{fig:neural_net}  
}
\killfloatstyle\capbtabbox[0.4\textwidth][]{%
\captionsetup{singlelinecheck = false, justification=raggedleft}
\raggedleft
\begin{tabular}{cc}
\hline
Parameter & Range \\ \hline
\# of spans & 1 to 60, step=2   \\ 
$L_\text{span}$ [km] & 10 to 120, step=1    \\ 
$\alpha$ [dB/km] & 0.19 to 0.275\\ 
Modulation & QPSK, 16QAM \\ 
$\Rsym$ & 35~GBd, 69~GBd \\ 
Data rate & 100G, 200G \\ 
\Ptx{} [dBm] & -6 to 2.5, step=0.5 \\ \hline
\end{tabular}
}{%
  \caption{Data generation parameters}
  \label{tab:in_params}
}
\end{floatrow}
\end{figure}


The choice of parameter space is based on representing the specifications and capabilities of real networks and up-to-date equipment while being sufficiently broad, such that the resulting ML model is versatile and can be applied to various configurations. Hence, we consider dispersion-uncompensated links using EDFAs at the end of each span, potentially supported by Raman in-line amplification (ILA) for high-loss spans. Restricting ourselves to standard single mode fiber of fixed dispersion ($D$=16.7ps/nm/km) and nonlinearity ($\gamma$=1.3 1/W/km), several fiber and system parameters were varied uniformly over a wide range, as listed in Table~\ref{tab:in_params}. Data channels that carry 100G (QPSK at 35GBd) or 200G (QPSK at 69GBd or 16QAM at 35GBd) are generated, with the WDM spectral width adjusted to either 50 or 75 GHz, depending on the symbol rate. The channel configurations reflect a subset of configurations available with ADVA's TeraFlex transponder \cite{adva21}. The spectral occupancy over the entire C-band was selected randomly between 75\% to 95\%. The transmit power of the channels was varied for each channel as per Table~\ref{tab:in_params}. For further randomization, the channel under test (CUT) was uniformly chosen between all generated channels. Evaluating the NLI for one setting constitutes one of the 130,000 generated data points.

\vspace{-7pt}
\subsection{Specifications of the ML Model}
\vspace{-5pt}
Two different feed-forward neural networks (NNs) were employed for QoT estimation. The baseline is a standard ANN using the leaky rectified linear unit (ReLU) activation function. This ANN is compared to an ANN using the scaled exponential linear unit (SeLU) activation function, also called self-normalizing neural network (SNN), which has been shown to outperform other NNs on a variety of data sets \cite{SNN}. Preliminary tests suggested the use of a cone-shaped ANN, in which the number of neurons in each layer are successively halved from the first hidden layer to the output layer. For the SNN, a constant number of neurons was used. Both models were trained for 50 epochs, minimizing the RMS error with the adamax optimizer. We used a batch size of 64 and started with a learning rate of 0.01, divided by 10 every 10th epoch. A grid search was used to determine the number of hidden layers and the number of neurons for the ANN and the SNN. The NN architecture and the used input parameters are shown in Fig.~\ref{fig:neural_net}. The choice of input parameters was found to be crucial for the performance of the ML model. While a large input space potentially allows for a finer regression model, the required size of the data set for training becomes prohibitively large. We heuristically determined a mix of scalars and averaged metrics that have a strong influence on NLI. For example, instead of using up to 60 individual span lengths as input, we found that the link architecture is well represented by using the average, min., max., variance, and the average of the cumulative sum. For the grid we use specific parameters (symbol rate $\Rsym$, transmit power and distance to CUT) for the four closest neighbors to the CUT. Additionally we use the number of channels $N_\text{ch}$ in the 10 closest 150~GHz areas around the CUT center frequency and further parameters shown in Fig.~\ref{fig:neural_net}, representing the full grid. The input data is normalized. The best performance for the ANN was achieved with 8 hidden layers, starting with 512 neurons on the first hidden layers, and for the SNN with 16 layers with 64 neurons each. For training and evaluation, we followed the usual 70/15/15 split of the data into training, validation and test set, respectively.


\vspace{-7pt}
\subsection{Numerical Results Based on Synthetic Simulation Data}
\vspace{-5pt}
We verify the performance of the ML models on a test set of $20,000$ datapoints using the EGN model results as baseline and the SNR deviation $\Delta \text{SNR} = |\text{SNR}_\text{ML}-\text{SNR}_\text{EGN}|$ as metric. Both ML models gave a mean SNR deviation of 0.07~dB on the synthetic data. The SNN has a slightly lower maximum SNR deviation of 0.58~dB compared to 0.64~dB for the ANN. Both models predict the NLI for the entire test set in under 1s (i7-7500U CPU with $12$GB RAM), thus having a prediction time in the order of microseconds per test case, which is orders of magnitude faster than the full EGN model \cite{EGN} requiring on average more than 1 minute for a single datapoint.

\vspace{-9pt}
\section{Live Network Study}\label{sec:results}
\vspace{-7pt}
\subsection{System and Link Configuration}
\begin{figure}[t]
\begin{center}
\includegraphics[width=\textwidth]{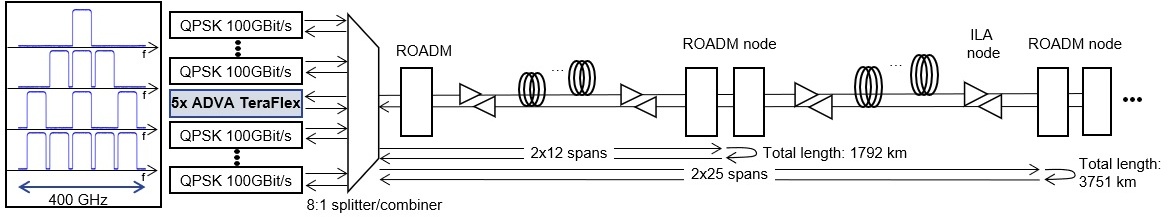}
\caption{Live network link with two loopback locations for the evaluated ADVA TeraFlex channels.}
\label{fig:setup}
\end{center}
\vspace*{-1.2\baselineskip}
\end{figure}

\vspace{-5pt}
The ML-based QoT estimation is performed in a commercial pan-European live network operated by Tele2 Estonia spanning several thousand kilometers and has the majority of the C-band filled with high-margin 100G QPSK channels. In addition to the live channels carrying production traffic, five ADVA TeraFlex transponders were installed for testing purposes. Our test-channels were inserted into a dedicated add/drop port of the terminal ROADM using an 8-port splitter/combiner module and occupied 400~GHz spectrum around 193.95 THz inside the C-band, configured as an optical spectrum-as-a-service. Two different loopback locations allowed measurements on a single site and gave total link lengths of 1792~km and 3751~km. Derived from the BER, the effective SNR, sometimes called generalized SNR (GSNR), of the center channel was recorded. The used configurations were 100G and 200G QPSK on the long link, and 200G QPSK and 16QAM on the short link. For each CUT configuration, the four WDM neighbours, having the same modulation format, were varied according to the on/off modes shown on the left of Fig.~\ref{fig:setup}. Additionally the on/off modes were repeated for 200G QPSK neighbors in the case of 200G 16QAM and 100G QPSK CUTs, leading to a total of 22 different channel plans.

%
%
This allows to capture the NLI impact of neighbouring channels on the CUT and thus to efficiently test the model's capability of QoT estimation. The linear SNR is calculated offline, including penalty terms for in-line ROADM filtering of the 69 GBd channels ($\sim2$ dB) and for any overlap of the 69 GBd channels on a 75 GHz grid (0.2 dB). Finally, the deviation from the measured values to the EGN and ML models are computed.

\vspace{-7pt}
\subsection{Numerical Results}
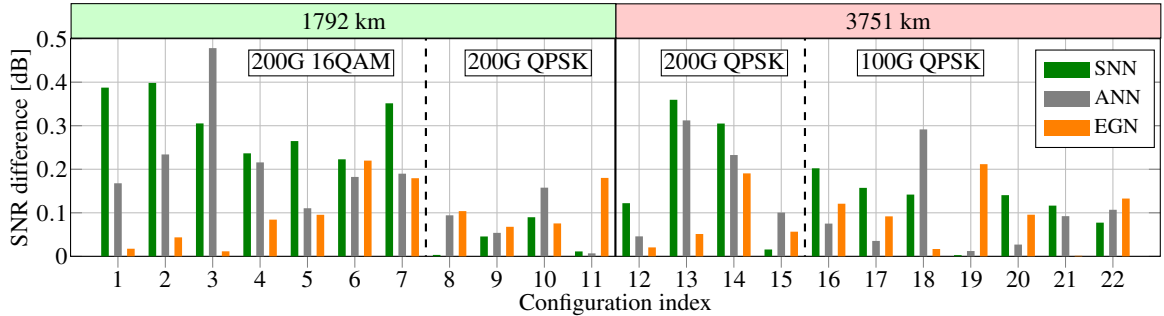
\begin{figure}[t]
\begin{center}
  \begin{tikzpicture}[font=\small]

\begin{axis}[%
ybar,
width=1\textwidth,
height=\FigureHeight,
bar width=0.1cm,
xmin=0,
xmax=23,
xlabel={Configuration index},
ymin=0,
ymax=0.5,
ytick={0,0.1,0.2,0.3,0.4,0.5},
ylabel={SNR difference [dB]},
xmajorgrids,
ymajorgrids,
legend style={at={(0.995,0.95)}, anchor=north east},
xmajorgrids,
xtick=data,
inner sep=1pt,
legend image code/.code={%
      \draw[#1,draw=none] (0cm,-0.1cm) rectangle (0.6cm,0.1cm);
    }   
]

\addplot[draw=none, fill=green!50!black,solid]
table[row sep=crcr] {%
1	0.387486013 \\ 
2	0.398317318 \\ 
3	0.305289895 \\ 
4	0.236571484 \\ 
5	0.264740704 \\ 
6	0.222836058 \\ 
7	0.351466044 \\ 
8	0.003261008 \\ 
9	0.045624641 \\ 
10	0.089806567 \\ 
11	0.011271213 \\ 
12	0.122076997 \\ 
13	0.359565991 \\ 
14	0.30500275 \\ 
15	0.015712754 \\ 
16	0.202170358 \\ 
17	0.15727678 \\ 
18	0.141854981 \\ 
19	0.002785316 \\ 
20	0.140511737 \\ 
21	0.116493856 \\ 
22	0.077383863 \\ 
};
\addlegendentry{SNN};

\addplot[draw=none, fill=black!50,solid]
table[row sep=crcr] {%
1	0.167790451 \\ 
2	0.234012088 \\ 
3	0.478173824 \\ 
4	0.2157583 \\ 
5	0.110443862 \\ 
6	0.182370421 \\ 
7	0.189837151 \\ 
8	0.094219331 \\ 
9	0.053937559 \\ 
10	0.157732211 \\ 
11	0.006850122 \\ 
12	0.045867726 \\ 
13	0.312209545 \\ 
14	0.232695649 \\ 
15	0.100405596 \\ 
16	0.075207053 \\ 
17	0.035411837 \\ 
18	0.291483406 \\ 
19	0.012199562 \\ 
20	0.027025807 \\ 
21	0.092271756 \\ 
22	0.106931848 \\ 
};
\addlegendentry{ANN};

\addplot[draw=none, fill=orange,solid]
table[row sep=crcr] {%
1	0.017483034 \\ 
2	0.043599586 \\ 
3	0.011522074 \\ 
4	0.084293882 \\ 
5	0.095425255 \\ 
6	0.219852512 \\ 
7	0.179443188 \\ 
8	0.10389084 \\ 
9	0.067855975 \\ 
10	0.075594398 \\ 
11	0.180133797 \\ 
12	0.020518066 \\ 
13	0.051329868 \\ 
14	0.190528744 \\ 
15	0.056619148 \\ 
16	0.120859976 \\ 
17	0.091822088 \\ 
18	0.016854446 \\ 
19	0.211690755 \\ 
20	0.09563322 \\ 
21	0.001318645 \\ 
22 	0.132739318 \\ 
};
\addlegendentry{EGN};

\draw[thick] (rel axis cs:0.5,1) -- (rel axis cs:0.5,0);

\draw[dashed,thick] (7.5,1.5) -- (7.5,0);

\coordinate (rect1_left) at (rel axis cs:0,1);
\coordinate (rect2_left) at (rel axis cs:0.5,1.15);

\coordinate (rect1_right) at (rel axis cs:1,1.15);
\coordinate (rect2_right) at (rel axis cs:0.5,1);

\node[draw,inner sep=1pt,fill=white,anchor=north,align=center] at (rel axis cs:0.23,0.95) {200G 16QAM};
\node[draw,inner sep=1pt,fill=white,anchor=north,align=center] at (rel axis cs:0.42,0.95) {200G QPSK};
\draw[dashed,thick] (15.5,1.5) -- (15.5,0);
\node[draw,inner sep=1pt,fill=white,anchor=north,align=center] at (rel axis cs:0.6,0.95) {200G QPSK};
\node[draw,inner sep=1pt,fill=white,anchor=north,align=center] at (rel axis cs:0.78,0.95) {100G QPSK};

\end{axis}%

\node[draw, anchor=south west, fill=green!20!white, rectangle, from={rect1_left to rect2_left}] {1792~km};
\node[draw, anchor=north east, fill=red!20!white, rectangle, from={rect2_right to rect1_right}] {3751~km};
\end{tikzpicture}%

\captionsetup{width=5.5in}
\caption{SNR deviation between live network measurement and the ANN, SNN, and EGN models for 22 different channel plans. The boxes give the data rate and modulation format of the CUT.}
\label{fig:Delta_SNR}
\end{center}
\vspace*{-1.4\baselineskip}
\end{figure}

\vspace{-5pt}
The performance of the ML models and the EGN model on the field measurement data is shown in  Fig.~\ref{fig:Delta_SNR} displaying the SNR deviations for all 22 channel plans. A maximum SNR deviation of 0.22, 0.40, 0.48 dB and a mean deviation of 0.1, 0.18, 0.15 dB is observed for the EGN, SNN and ANN model, respectively. At less than 0.5 dB SNR deviation, the SNR of the ML models is computed within microseconds, which is several orders of magnitude faster than the EGN model. The two NNs show similar performance, with the SNN having a slightly lower maximum SNR deviation, but a slightly higher mean deviation than the ANN. Among the excellent fit on all configurations, the highest deviations of both ML models are observed for the 200G 16QAM channel. This can be attributed to the CUT's modulation format not being included in the model's input parameters. As including this parameter into the current data set leads to a disproportional deterioration of the performance of the QPSK channels, this requires a modified and extended data set, which is left for further research.

\vspace{-9pt}
\section{Conclusions}
\vspace{-7pt}
We have introduced an ML model for QoT estimation of unestablished lightpaths in a live production network. The model is trained on a synthetic database populated by the EGN data and uses an input parameter space that models the NLI of a lightpath well without becoming prohibitively large. In combination with a deep network architecture, this enables to transfer the model performance to a live network, showing a maximum SNR error of less than 0.5 dB compared to the field measurements. The model computes the NLI of a lightpath within microseconds, demonstrating its potential for real-time network management and operation applications.
\\[-1pt]
{\scriptsize
The work has been partially funded by the German Ministry of Education and Research in the project OptiCON (\textbf{\#16KIS0989K}).}



\vspace{-12pt}
\begin{thebibliography}{99}
\vspace{-6pt}
\footnotesize
\setlength{\itemindent}{-12pt}%

\bibitem{EGN} A. Carena \textit{et al.}, ``EGN Model of Nonlinear Fiber Propagation,'' Opt. Express, \textbf{22}(13), 2014.
\bibitem{Morais} R. M. Morais \textit{et al.}, ``Machine Learning Models for Estimating Quality of Transmission in DWDM Networks,'' JOCN, \textbf{10}(10), 2018.
\bibitem{Aladin} S. Aladin \textit{et al.}, ``Quality of Transmission Estimation and Short-Term Performance Forecast of Lightpaths,'' JLT, \textbf{38}(10), 2020.
\bibitem{Rottondi} C. Rottondi \textit{et al.}, ``Machine-Learning Method for Quality of Transmission Prediction of Unestablished Lightpaths,'' JLT, \textbf{10}(2), 2020.
\bibitem{Gao} Z. Gao \textit{et al.}, ``ANN-Based Multi-Channel QoT-Prediction over a 563.4-KM Field-Trial Testbed,''JLT, \textbf{38}(9), 2020.
\bibitem{adva21} ADVA, ``TeraFlex", www.adva.com/en/products/open-optical-transport/fsp-3000-open-terminals/teraflex, Accessed: 2021-01-23.
\bibitem{SNN} G. Klambauer \textit{et al.}, ``Self-Normalizing Neural Networks,'' arXiv:1706.02515, 2017.


\end{thebibliography}
\end{document}